\documentclass[FBSedit,FBSmath,ecsub]{FBSsuppl}
\usepackage{graphicx}
\usepackage{amssymb}
\begin{document}
\title{Nucleon Electromagnetic and Axial Form Factors in Point-Form
Relativistic Quantum Mechanics}
\author{R.F.~Wagenbrunn\instnr{1}, S.~Boffi\instnr{2},
L.Ya.~Glozman\instnr{1}, W.~Klink\instnr{3},
W.~Plessas\instnr{1}, M.~Radici\instnr{2}
}                     
\instlist{Institut f\"ur Theoretische Physik, Universit\"at Graz, 
A-8010 Graz, Austria \and
Dipartimento di Fisica Nucleare e Teorica, Universit\`a di Pavia,
and Istituto Nazionale di Fisica Nucleare, Sez. di Pavia,
Via Bassi 6, I-27100 Pavia, Italy \and
Department of Physics and Astronomy, University of Iowa,
Iowa City, IA 52242, USA}
\maketitle
\begin{abstract}
Results for the proton and neutron electric and magnetic form factors as
well as the nucleon axial form factor are presented for 
constituent quark models, based on either one-gluon-exchange and 
Goldstone-boson-exchange dynamics. The calculations are performed in a
covariant framework using the point-form approach to relativistic quantum
mechanics. The only input to the calculations is the nucleon wave function
of the corresponding constituent quark model. A comparison is given to 
results of the instanton-induced constituent quark model treated 
with the Bethe-Salpeter equation.
\end{abstract}
It has become fairly obvious that nucleon electromagnetic as well as 
axial form factors need a relativistic treatment. In order to satisfy 
all requirements of Lorentz covariance one can either proceed along 
relativistic field theory or relativistic (Poincar\'e-invariant) 
quantum mechanics. The latter approach appears promising whenever one 
can deal with a fixed number of particles (or a finite number of 
degrees of freedom). Usually one follows one of the three forms of 
dynamics, where the generators of the Poincar\'e group are minimally 
affected by interactions. Dirac defined them as instant, front, and 
point forms \cite{Dirac:1949cp}.

The point form is very convenient for practical calculations since
only the components of the four-momentum operator $\hat{P}^{\mu}$
contain interactions. Therefore all Lorentz transformations remain
purely kinematic and the theory is manifestly covariant (for more 
details see ref. \cite{Klink:1998hb}). In order
to construct the interacting four-momentum operators one follows the 
Bakamjian-Thomas construction \cite{Bakamjian:1953kh}, where the
interaction is introduced into the mass operator $\hat{M}$
by adding to the free mass operator $\hat{M}_{\rm fr}$ the interacting
term $\hat{M}_{int}$. Through multiplication with the free four-velocity
operator $\hat{V}_{\rm fr}^{\mu}$, which is not affected by interactions,
one obtains
\begin{equation}
\hat{P}^{\mu} =
\hat{M}\hat{V}_{\rm fr}^{\mu} 
= (\hat{M}_{\rm fr}+\hat{M}_{int})\hat{V}_{\rm fr}^{\mu}.
\end{equation}
Poincar\'e invariance implies that $\hat{M}$ commutes with
$\hat{V}_{\rm fr}^{\mu}$ and is a scalar under Lorentz transformations.
Therefore eigenstates of the four-momentum operator are simultaneous
eigenstates of both the mass and the velocity operators. As a consequence
the motion of the system as a whole and the internal motion are separated.
The latter is described by a wave function containing only the internal
degrees of freedom. It can be found by solving the eigenvalue problem for
the mass operator 
\begin{equation}
\hat{M}\Psi = M\Psi.
\end{equation}
\begin{figure}
\parbox{0.5\hsize}{
\begin{center}
\includegraphics[width=1.02\hsize,bb=35 14 339 253,clip=]{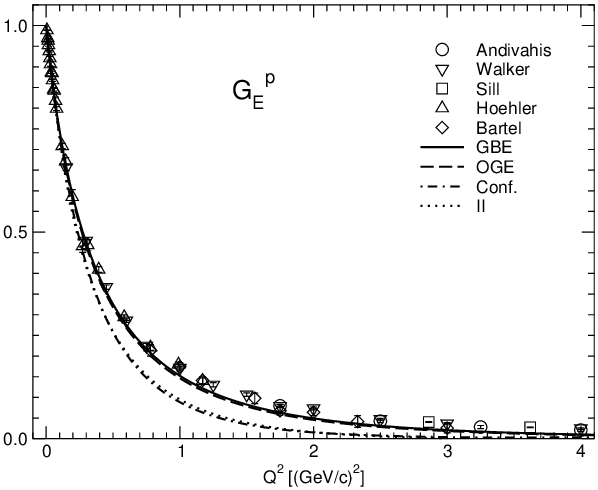}
\vspace{2mm}

\includegraphics[width=1.02\hsize,bb=35 14 339 253,clip=]{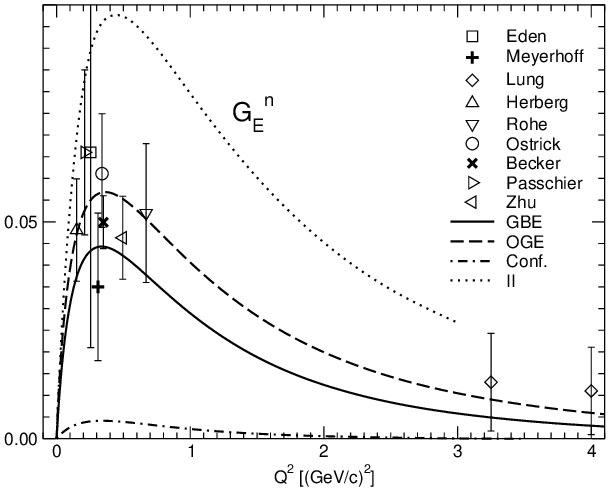}
\vspace{2mm}

\includegraphics[width=1.02\hsize,bb=35 14 339 253,clip=]{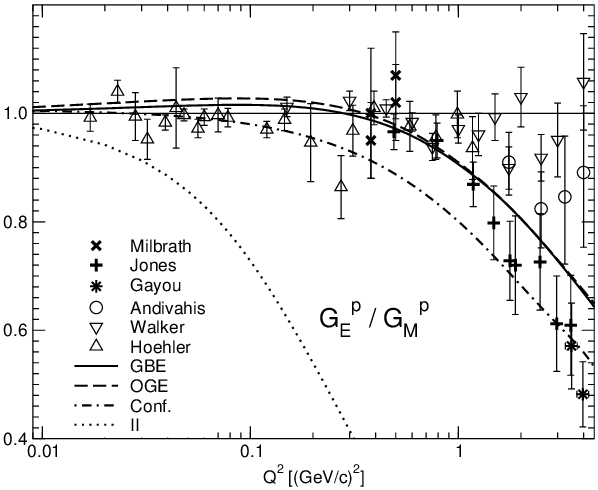}
\end{center}}
\parbox{0.5\hsize}{
\begin{center}
\includegraphics[width=1.02\hsize,bb=35 14 339 253,clip=]{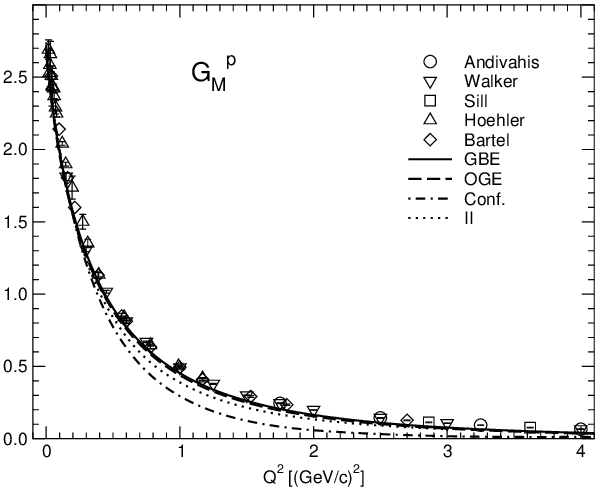}
\vspace{2mm}

\includegraphics[width=1.02\hsize,bb=35 14 339 253,clip=]{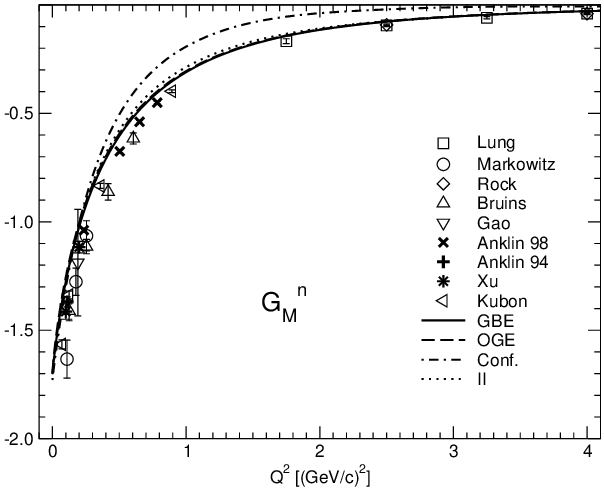}
\vspace{2mm}

\includegraphics[width=1.02\hsize,bb=35 14 339 253,clip=]{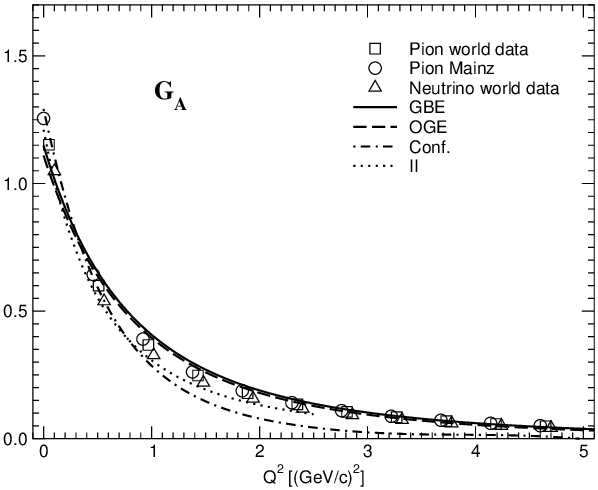}
\end{center}}
\vspace*{-0.2cm}
\caption{Predictions of different CQMs for the nucleon electromagnetic and
axial form factors. The solid and dashed lines represent our PFSA
results for the GBE and OGE CQMs, respectively; the dash-dotted line
refers to the case with confinement only. The dotted lines show the
results of the II CQM within the Bethe-Salpeter approach after
ref. \cite{Merten:2002nz}.}
\end{figure}
\vspace*{-1mm}

The electromagnetic and axial form factors of the nucleons are 
obtained by sandwiching the electromagnetic and axial current operators
between eigenstates $\left|P,\Sigma\right\rangle$, where $P$ and 
$\Sigma$ are the eigenvalues of the total momentum and the 
z-component of the total angular momentum (for details see
refs. \cite{Wagenbrunn:2000es}).
The calculation boils down to matrix elements of the current
operators between free three-particle states
$\left|p_1,p_2,p_3;\sigma_1,\sigma_2,\sigma_3\right\rangle$, where
$p_i$ are the individual quark four-momenta and $\sigma_i$ their
spin projections. At the present stage we cannot yet
deal with the full current operators but have to truncate them
to one-body operators by the so-called point-form spectator
approximation (PFSA) \cite{Klink:1998hb}
\begin{equation}
\begin{array}{l}
\left\langle p'_1,p'_2,p'_3;\sigma'_1,\sigma'_2,\sigma'_3\right|
\hat{J}^\mu(0)
\left|p_1,p_2,p_3;\sigma_1,\sigma_2,\sigma_3\right\rangle
=\\[1ex]
2E_2\delta(\vec{p}'_2-\vec{p}_2)\delta_{\sigma_2'\sigma_2}
2E_3\delta(\vec{p}'_3-\vec{p}_3)\delta_{\sigma_3'\sigma_3}
\left\langle p'_1,\sigma'_1\right|
\hat{j}^\mu(0)
\left|p_1,\sigma_1\right\rangle,
\end{array}
\end{equation}
and similarly for the axial current $\hat{\vec{A}}{}^\mu(0)$ with
$\hat{j}^\mu(0)$ replaced by $\hat{\vec{a}}{}^\mu(0)$. By the small
letters we indicate free one-body currents of the constituent quarks.
The axial current is denoted as a vector in isospin space. With 
regard to the PFSA it is important to notice that the impulse
$\tilde{q}=p_1'-p_1$ delivered to the single constituent quark is
different from the impulse $q=P'-P$ delivered to the nucleon as a whole.
The momentum transfer $\tilde{q}$ can be uniquely determined
from $q$ and the two spectator conditions $\vec{p}\,'_2=\vec{p}_2$ and
$\vec{p}\,'_3=\vec{p}_3$. For the one-body current matrix elements in 
the above equation we employ the usual expressions for electromagnetic
and axial currents of pointlike spin-$\frac{1}{2}$ particles (see refs.
\cite{Wagenbrunn:2000es}).

The results for all electromagnetic and axial form factors from the
Goldstone-boson-exchange (GBE) CQM of ref. \cite{Glozman:1998ag}
have already been published in refs.
\cite{Wagenbrunn:2000es}. There it was found that relativistic effects 
are most important. The direct predictions of the GBE CQM in PFSA come
remarkably close to the experimental data in all instances. This 
observation has recently been confirmed also with regard to the 
electric radii and magnetic moments not only of the nucleons but all 
(measured) octet and decuplet ground states \cite{Berger}.

Here we present a comparison of nucleon form factor results from 
different CQMs and different relativistic approaches. First, we 
compare the PFSA predictions of the GBE CQM with analogous results 
from a CQM whose hyperfine interaction is based on one-gluon exchange
(OGE), namely a relativized version of the Bhaduri-Cohler-Nogami CQM 
\cite{BCN} as parametrized in ref. \cite{Theussl:2000sj}. Then we also 
provide a comparison to the results of an instanton-induced (II) CQM as  
obtained by the Bonn group within a Bethe-Salpeter (BS) formalism
\cite{Merten:2002nz}.

As is immediately evident from the results collected in Fig. 1,
the overall behaviour of the relativistic predictions appears quite 
reasonable in all cases. This confirms the previous findings that the 
inclusion of relativity is most important for the nucleon form 
factors. For their gross properties dynamical effects are of lesser 
relevance. Even a three-quark wave function that relies solely on
confinement produces the right 
features, except for the neutron. In this case a realistic wave 
function is required, with the mixed-symmetry spatial
components taken into account. For the proton form factors 
and also the nucleon axial form factor there is a striking similarity 
of the results obtained in PFSA and in the BS approach,
where the latter also uses a single-particle approximation for the
current operators. Only for the subtle details of the neutron electric 
form factor and the ratio of the proton electric to magnetic form 
factors the predictions of the II CQM fall short compared to 
experimental data. A comparison of the CQM results for electric radii 
and magnetic moments is given in ref. \cite{Berger}, yielding a 
picture congruent with the one found here.  
\vspace{-2mm}
\begin{acknowledge}
This work was supported by the Austrian Science Fund (Project P14806).
\end{acknowledge}


\begin{thebibliography}{1}

\expandafter\ifx\csname url\endcsname\relax
  \def\url#1{\texttt{#1}}\fi
\expandafter\ifx\csname urlprefix\endcsname\relax\def\urlprefix{URL }\fi

\bibitem{Dirac:1949cp}
P.~A.~M. Dirac: Rev. Mod. Phys. \textbf{21}, 392 (1949).

\bibitem{Klink:1998hb}
W.~H. Klink: Phys. Rev. C \textbf{58}, 3587 (1998).

\bibitem{Bakamjian:1953kh}
B.~Bakamjian and L.~H. Thomas: Phys. Rev. \textbf{92}, 1300 (1953).

\bibitem{Wagenbrunn:2000es}
R.~F. Wagenbrunn, S.~Boffi, W.~Klink, W.~Plessas, and M.~Radici: 
Phys. Lett. B \textbf{511}, 33 (2001);
L.~Ya. Glozman, M.~Radici, R.~F. Wagenbrunn, S.~Boffi, W.~Klink, and
W.~Plessas:
Phys. Lett. B \textbf{516}, 183 (2001);
S.~Boffi, L.~Ya. Glozman, W.~Klink, W.~Plessas, M.~Radici, and
R.~F. Wagenbrunn:
Eur. Phys. J. A \textbf{14}, 17 (2002).

\bibitem{Glozman:1998ag}
L.~Ya. Glozman, W.~Plessas, K.~Varga, and R.~F. Wagenbrunn:
Phys. Rev. D \textbf{58}, 094030 (1998).

\bibitem{Berger}
K.~Berger: Diploma Thesis, Univ. Graz (2002); K. Berger, W. Plessas, 
and R.F. Wagenbrunn: Prog. Part. Nucl. Phys., to appear; see also
K.~Berger, W.~Plessas, and R.F.~Wagenbrunn: These Proceedings.

\bibitem{BCN} R.K. Bhaduri, L.E. Cohler, and Y. Nogami,
Nuovo Cim. A \textbf{65}, 376 (1981).

\bibitem{Theussl:2000sj}
L.~Theu{\ss}l, R.~F. Wagenbrunn, B.~Desplanques, and W.~Plessas: 
Eur. Phys. J. A \textbf{12}, 91 (2001).

\bibitem{Merten:2002nz}
D.~Merten, U.~L{\"o}ring, K.~Kretzschmar, B.~Metsch, and H.~R.~Petry:
Eur. Phys. J. A \textbf{14}, 477 (2002).

\end{thebibliography}
\end{document}